\documentclass[%
 reprint,
nofootinbib,
 amsmath,amssymb,
 aps,
]{revtex4-2}

\usepackage{xcolor}

\usepackage{slashed}
\usepackage{graphicx}
\usepackage{dcolumn}
\usepackage{bm}
\usepackage{hyperref}

\begin{document}

\preprint{APS/123-QED}

\title{The Non-triviality of Dynamical Chern-Simons Gravity and the Standard Model }

\author{Stephon Alexander}
 \email{stephon\_alexander@brown.edu}
\author{Heliudson Bernardo}%
 \email{heliudson\_bernardo@brown.edu}
\affiliation{Brown Theoretical Physics Center and Department of Physics, Brown University,\protect\\  Barus Building, 340 Brook Street, Providence, RI 02912, USA}


\author{Cyril Creque-Sarbinowski}
 \email{ccreque@flatironinstitute.org}
\affiliation{Center for Computational Astrophysics, Flatiron Institute \protect \\ 162 5th Avenue, New York, New York 10010, USA}%


\date{\today}

\begin{abstract}
Given the growing interest in gravitational-wave and cosmological parity-violating effects in dynamical Chern-Simons (dCS) gravity, it is crucial to investigate whether the scalar-gravitational Pontryagin term in dCS persists when formulated in the context of the $\text{U(1)}_{\text{B}-\text{L}}$ anomaly in the Standard Model (SM). In particular, it has been argued that dCS gravity can be reduced to Einstein gravity after ``rotating away" the gravitational-Pontryagin coupling into the phase of the Weinberg operator -- analogous to the rotation of the axion zero-mode into the quark mass matrix. We find that dCS is nontrivial if the scalar field $\phi$ has significant space-time dependence from dynamics. We provide a comprehensive consideration of the dCS classical and quantum symmetries relevant for embedding a dCS sector in the SM. We find that, because of the B-L chiral gravitational anomaly, the scalar-Pontryagin term cannot be absorbed by a field redefinition. Assuming a minimal extension of the SM, we also find that a coupling of the dCS scalar with right-handed neutrinos induces both the scalar-Pontryagin coupling and an axion-like phase in the dimension-five Weinberg operator. We comment on the issue of gauging the  $\text{U(1)}_{\text{B}-\text{L}}$, the observational effects with these two operators present for upcoming experiments, and the origin of dCS gravity in string theory. 


\end{abstract}

\maketitle

\tableofcontents

\section{Introduction}
\label{sec:intro}

Given that the electroweak sector of the Standard Model (SM) maximally violates parity, it is natural to investigate whether parity-violation occurs in sectors beyond the SM. In particular, the gravitational sector has been under such observational and theoretical investigation over the past several years~\cite{Alexander:2009tp, Molina:2010fb, Ali-Haimoud:2011zme, Yagi:2012vf, Okounkova:2017yby, Cai:2021uup, Doneva:2021dcc, Ng:2023jjt}. The primary phenomenological motivation is that parity-violating modifications of general relativity (GR) lead to a myriad of distinct astrophysical and cosmological observables. 

As originally demonstrated in Ref.~\cite{Lue:1998mq}, the cosmic microwave background (CMB) is sensitive to parity violation through its EB and TB angular power spectra. In fact, utilizing Galactic foreground emission as a polarization-angle calibrator~\cite{Minami:2019ruj, Minami:2020xfg, Minami:2020fin}, there are recent hints that Planck's EB angular power spectrum shows such a signature \cite{Minami:2020odp, Eskilt:2022cff}.  In addition, there are further indications that the four-point galaxy correlation function is parity violating \cite{Philcox:2022hkh, Hou:2022wfj}, although the CMB's four-point correlators yield no analogous signal \cite{Philcox:2023ffy, Philcox:2023xxk}. Finally, it is possible that parity-violation may be encoded in the large-scale correlation of galaxy spins, as explored in Refs.~\cite{Lee:1999ii, Crittenden:2000wi, Yu:2019bsd, Motloch:2021mfz}. Clearly, as such parity-sensitive data accumulates, the need for well-motivated parity-breaking GR theories increases.

The simplest and most theoretically motivated effective theory of gravity that encodes parity violation is dynamical Chern-Simons (dCS) gravity\footnote{An earlier non-dynamical version that violates the strong equivalence principle of Chern-Simons gravity was first put forth in Ref.~\cite{Jackiw:2003pm}. See also \cite{Bellucci:1988ff, Bellucci:1990fa, Gates:1991qn} and references therein for works on the Lorentz CS form in supersymmetric theories.}There are different motivations to study such a theory. The most basic one is the effective theory perspective, which states one could add all possible terms compatible with symmetries. Thus, dCS gravity is a valid candidate for a classical scalar-tensor modification of GR. This approach to dCS has been explored in the literature before (see Ref.~\cite{Alexander:2009tp} and references therein).

DCS gravity was the framework within which the cosmic baryon/lepton number (B-L) asymmetry was instantiated in inflationary models through the production of parity asymmetry of chiral gravitational waves which sourced a B-L asymmetry by the end of inflation \cite{Alexander:2004us}. It was later demonstrated that these gravitational waves can induce a scalar four-point (trispectrum) that contributes to the primordial large-scale parity violation in galactic distributions~\cite{Creque-Sarbinowski:2023wmb}. In the strong gravity regime, such as compact binary systems, dCS gravity leads to distinct waveforms in the propagation and sourcing of gravitational waves as well as the phenomena of scalarization~\cite{Yunes:2009hc, Alexander:2021ssr}.

It is therefore important to consider how dCS connects with the standard model and its logical extensions. From the perspective of modified theories of gravity, such as $f(R)$ theories, the principle of general covariance is enough to formulate it. This is the case of versions of dCS that emerge as an effective theory of $4D$ Heterotic string theory, and there is no need to relate it directly to the standard model.  However, dCS gravity has a term where a pseudoscalar couples to the gravitational Chern-Simons form, which also encodes the B-L global anomaly\footnote{We shall call a classical symmetry anomalous if the path integral measure is not invariant under the symmetry transformation.} in the standard model. This structure has a close semblance to the axion coupling to the Yang-Mills Pontryagin density. For the latter case, in the presence of massless quarks and when the axion is stabilized at a value $\theta$, a field redefinition of the quark fields can eliminate the $\theta F\Tilde{F}$ coupling, rendering the CP-violating effects ignorable. In this paper, we investigate to what extent and how these axion features apply to the dCS pseudoscalar. This is relevant for any SM extension in which the dCS scalar-Pontryagin coupling appears after integrating out heavy fermions.

The paper is organized as follows. In the next section, we discuss the global shift symmetry of dCS gravity and use analogous axion results to comment on what to expect in the quantum gravitational regime. In Section \ref{sec:coupling_with_fermions}, we show that dCS gravity is non-trivial even after including fermions, highlighting some features relevant to the coupling with SM fermions. In Section \ref{sec:coupling_with_SM}, we propose an extension of the SM where a dCS sector appears after integrating out right-handed neutrinos. We conclude in Section \ref{sec:conclusion} with a discussion on the combined shift and B-L symmetry of the model and its gauging. Some aspects of the baryonic and leptonic number anomalies in the SM are summarized in Appendix \ref{SM_anomalies}. 

{\it Conventions}: We set $\hbar = c = 1$. Moreover, we work with the mostly plus metric-signature convention and the Weyl (chiral) basis for the Dirac matrices. 

\section{dCS Gravity and Axions}

\subsection{Continuous Shift Symmetry}

The dCS action is 
\begin{align}\label{dCS_action}
    S_{\text{dCS}}[g_{\mu\nu}, \phi] = \int d^4x \sqrt{-g}&\left(\frac{1}{2\kappa^2}R -\frac{1}{2}\partial_\mu \phi \partial^\mu \phi - \right.\nonumber\\
    &\left.-\lambda \phi \Tilde{R}_{\mu\nu\rho\sigma}R^{\mu\nu\rho\sigma}\right),
\end{align}
where $g_{\mu\nu}$ is the spacetime metric with $g$ its determinant, $\phi$ is the dCS pseudoscalar, $R$ is the Ricci scalar, $R^{\mu\nu\rho\sigma}$ is the Riemann tensor with $\Tilde{R}_{\mu\nu\rho\sigma}$ its Hodge dual, and $\lambda$ the dCS coupling constant. 

Under the global symmetry $\phi(x^\mu) \to \phi(x^\mu) + \chi$, Eq.~\eqref{dCS_action} changes by a constant shift  since the last term can be written as the derivative of the gravitational Chern-Simons (CS) current $K^\mu$,
\begin{align}
    \Tilde{R}_{\mu\nu\rho\sigma} R^{\mu\nu\rho\sigma} &= 2\epsilon^{\mu\nu\rho\sigma}\partial_\mu\left[\text{tr}\left(\omega_{[\nu} \partial_\rho \omega_{\sigma]} +\frac{2}{3}\omega_{[\nu} \omega_\rho \omega_{\sigma]} \right)\right]\nonumber\\
    &\equiv \nabla_\mu K^\mu, 
\end{align}
where $\epsilon^{\mu\nu\rho\sigma}$ are the contravariant components of the Levi-Civita tensor, and the trace is taken over the indices of the Lorentz-algebra valued spin connection ${\omega_{\mu}}^{a}_{\;\;b}$. The equations of motion will be invariant under constant scalar shifts. We will call this shift symmetry $\text{U}(1)_{\text{dCS}}$. Its associated Noether current is
\begin{equation}
    j^\mu_{\text{dCS}} = -\partial^\mu \phi +\lambda K^\mu,
\end{equation}
which satisfy
\begin{equation}\label{div_dCScurrent}
    \nabla_\mu j^\mu_{\text{dCS}} = 0
\end{equation}
upon using $\phi$'s equation of motion $\Box \phi - \lambda \Tilde{R}R =0$.\footnote{We note that the $S_{\text{dCS}}$ is not strictly invariant under global $\phi$ shifts because the dCS Lagrangian changes by a total derivative. Symmetries satisfying this property are often referred to as quasi-symmetries, so a $\text{U}_\text{dCS}(1)$ transformation is more properly referred to as a quasi-symmetry of the action, although we will make no such distinction in what follows.}

For field theories in a flat spacetime with vanishing boundary conditions, Eq.~\eqref{div_dCScurrent} guarantees the existence of a conserved charge. In curved spacetime, a conserved Noether charge is obtained if we assume a globally hyperbolic and asymptotically flat spacetime. In this case, we then integrate Eq.~\eqref{div_dCScurrent} over a spacetime volume $V$ foliated by spacelike surfaces $\Sigma_t$ whose boundaries $\partial \Sigma_t$ are 2-spheres with a very large radius (compared to the maximum curvature scale inside $V$), where $t$ is the parameter of the timelike curves perpendicular to $\Sigma_t$. We have
\begin{align}\label{int_Noether_current}
    0&= \int_V d^4 x \sqrt{-g}\nabla_\mu j^\mu = \int_{\partial V} d^3x \sqrt{|h|} n_\mu j^\mu \nonumber\\
    &= \int_{\Sigma_{t_1}} dS j^0 - \int_{\Sigma_{t_2}} dS j^0 + \int_{\mathbb{R}\times S^2} dS n_i j^i.
\end{align}
Note that $j^\mu$ is defined up to a term $\Delta j^\mu$ that vanishes at the boundary of $V$. If the last integral in the second line of Eq.~\eqref{int_Noether_current} vanishes, then we have that
\begin{equation}
    Q = \int_{\Sigma_t} dS j^0
\end{equation}
is independent of $\Sigma_t$ and thus is conserved. Typically it is assumed that $n_i j^i \to 0$ asymptotically such that the last integral in Eq.~\eqref{int_Noether_current} vanishes and we have charge conservation. 

Let us now apply these facts to dCS gravity. Assuming that $\partial^i \phi$ vanishes at the boundary, only the gravitational CS current will contribute to the integral of $n_ij^i$. However, the spin-connection will be pure gauge on the cylinder at infinity, and we can choose a gauge where its time component $\omega_0$ vanishes there \cite{Peccei:1977np}. In such a case, the CS contribution to $n_ij^i_\text{dCS}$ is null. Then, the conserved charge is 
\begin{equation}\label{dCScharge}
    Q_\text{dCS} = \int_{\Sigma_t} dS \left(\Dot{\phi} + \lambda K^0\right),
\end{equation}
This charge is precisely the momentum of $\phi$, with the contribution $K^0$ coming from the fact that the Lagrangian contains a term proportional to $\lambda\Dot{\phi}K^0$. If $Q_\text{dCS}$ were gauge invariant, the expression above would be independent of the gauge choice $\omega_0|_{\partial\Sigma_t}=0$. As shown below, $Q_\text{dCS}$ changes under large Lorentz gauge transformations. Under a Lorentz transformation $\Lambda$, we have that the gravitational CS current transforms as $K^\mu \to K'^\mu$, where
\begin{align}
    K'^\mu = K^\mu - 2\epsilon^{\mu\nu\rho\sigma}\text{tr}&\left[\partial_\nu\left(\partial_\rho \Lambda \Lambda^{-1} \omega_\sigma\right) + \right.\nonumber\\
    &\left.+\frac{1}{3}\partial_\nu \Lambda \Lambda^{-1} \partial_\rho \Lambda \Lambda^{-1} \partial_\sigma \Lambda \Lambda^{-1}\right].
\end{align}
It follows that \cite{Jackiw:1976pf}
\begin{align}
    Q' = Q -\frac{2}{3}\lambda\int_{\Sigma_t} dS \epsilon^{0ijk} \text{tr}\left[\partial_i \Lambda \Lambda^{-1} \partial_j \Lambda \Lambda^{-1} \partial_k \Lambda \Lambda^{-1}\right] 
\end{align}
and thus 
\begin{equation}\label{Q_gauge_transf}
    Q'_\text{dCS} = Q_\text{dCS} - 16\pi^2 \lambda \nu(\Lambda),
\end{equation}
where $\nu(\bf \Lambda)$ is the winding number of the corresponding Lorentz transformation (that might be non-vanishing even with $\Lambda(x^i) \to \mathbf{1}$ asymptotically at infinity). In particular, we have that the Noether charge is only invariant under transformations such that $\nu(\Lambda) = 0$, i.e. for transformations continuously connected to the identity. But the charge transforms non-trivially for large Lorentz transformations. Eq.~\eqref{Q_gauge_transf} then shows that the charge $Q_\text{dCS}$ associated with global shifts in the dCS scalar transform non-trivially under large gauge transformation while still being conserved. 

Some of the above results are analogous to axions coupled to non-Abelian gauge fields \cite{Peccei:1977ur, Peccei:1977hh, Weinberg:1977ma, Wilczek:1977pj}. To better understand what can be applied to the gravitational case, we shall review some aspects of axions in the next section. 

\subsection{Non-Abelian Axions}

We wish to understand what happens with the shift symmetry and the conserved charge when we quantize the scalar $\phi$ in theory. Because all the results above have analogs for axions and their coupling to gauge fields, we will first consider the system
\begin{equation}
    S = \int d^4x \left(-\frac{1}{2}\partial_\mu a \partial^\mu a -  \lambda a \text{tr}\Tilde{F}_{\mu\nu}F^{\mu\nu}\right),
\end{equation}
which includes an axion-like field $a$ coupled to a SU$(2)$ gauge field $A_\mu^a$ via the term proportional to $\lambda$. We can consider this system in Minkowski space and Cartesian coordinates for simplicity since curvature is unimportant for the axion discussion in this section. Classically, the action above has the global shift symmetry $a\to a + c$ with conserved current
\begin{equation}
    j^\mu = -\partial^\mu a + \lambda C^\mu, 
\end{equation}
where $C^\mu$ is the Chern-Simons current of the gauge field
\begin{equation}
    C^\mu = 2\epsilon^{\mu\nu\rho\sigma}\text{tr}\left(A_\nu \partial_\rho A_\sigma +\frac{2}{3}A_\nu A_\rho A_\sigma\right),
\end{equation}
and the trace is now taken over the adjoint of the gauge group. The Noether charge is 
\begin{equation}
\label{dCSQ}
    Q_a = \int_\Sigma dS \left(\Dot{a} + \lambda  C^0\right),
\end{equation}
where we chose the gauge $A^0=0$ without loss of generality. Under a gauge transformation $A_\mu \to g^{-1} A_\mu g + g^{-1}\partial_\mu g$ where
\begin{equation}\label{gauge_group_elem}
    g(x^i) = e^{if(r)\hat{r}\cdot \tau}    
\end{equation}
with $\tau^a$ the generators of SU$(2)$ and $f(r)\to 2\pi n$ as $r\to \infty$, the charge $Q_a$ transforms as \cite{Jackiw:1976pf, Callan:1977gz}
\begin{equation}\label{Qa_gauge_transf}
    Q_a' = Q_a - 16\pi^2\lambda n(g).
\end{equation}

This gauge dependence does not introduce any issues because once we fix the background gauge field configuration, the system will not be invariant under large gauge transformations with a non-trivial winding number. To see that explicitly, note that a given background configuration $A^\mu(t, x^i)$ should approach a pure gauge one as $r\to \infty$, and we can choose it to be of the form \eqref{gauge_group_elem} with $f(r) \to 2\pi m$ at infinity. In other words, the homotopy class of the gauge configuration is part of the system's definition so that large gauge transformations would change the system altogether. Thus, by definition, large gauge transformations cannot be a symmetry of the system, and it is no surprise that such a gauge transformation changes $Q_a$.

The conservation of the homotopy-class dependent $Q_a$ implies that the dynamical evolution cannot change the winding number of the background configuration $A_{(\nu)}^\mu$. In fact, even if we promote $A^\mu$ to a dynamical field, there are no solutions to the equations of motion in Lorentzian signature that would allow a change in winding number \cite{Peccei:1977np}. This also guarantees that if $A^\mu$ is pure gauge asymptotically at $t\to \pm \infty$, the action is invariant under global $a$-shifts: we have 
\begin{equation}
    S[a+\chi] = S[a] - \lambda \chi \int d^4 x \text{tr}\Tilde{F} F,
\end{equation}
but 
\begin{align}
    \int d^4 x \text{tr}\Tilde{F} F &= \int d^4x \partial_\mu C^\mu = \left.\int d^3x C^0\right|_{t=-\infty}^{t= \infty} =\nonumber\\
    &=32\pi^2\left[n(t=+\infty) - n(t=-\infty)\right] = 0,
\end{align}
since the homotopy class of the gauge field does not change in time.

However, if $A_\mu^a$ is promoted to a dynamical field that should also be quantized, the argument in the paragraph above cannot be applied. If we also quantize the gauge field, we cannot exclude the possibility of quantum tunneling between gauge configurations of different homotopy classes. This is precisely what happens in quantum chromodynamics (QCD) and general non-Abelian gauge theories, as manifested by instanton solutions. This is the reason why the vacuum structure of gauge theories is non-trivial \cite{Callan:1977gz}. 

With a quantized gauge field, the system cannot be fixed in given homotopy class because there is always a probability for the gauge field to tunnel to configurations of different classes. In other words, the partition function includes a sum over the instantons connecting different homotopy classes,
\begin{align}
    Z = \sum_\nu \int &[d\mu_{(\nu)} dA_{(\nu)} da] \exp\left({i S_{\text{gauge}}}\right)\times\exp\left({i\theta \nu}\right)\\
    \times &\exp\left[{i\int \left(-\frac{1}{2}\text{tr}F^2 - \frac{1}{2}(\partial a)^2 - \lambda a \Tilde{F}F\right)}\right],\nonumber
\end{align}
where we wrote the (homotopically trivial) gauge fixing and Fadeev-Popov contributions in $S_{\text{gauge}}$ and $[d\mu_{(\nu)}]$. Then, the vacuum state of the whole system depends on $\theta$, but all physical observables would be the same regardless of the vacuum $\theta$ we use to compute them. In the quantum theory defined above, the specific value of theta is unphysical because of the shift symmetry at the quantum level: since
\begin{equation}
    \nu = \frac{1}{32\pi^2}\int d^4 x \Tilde{F} F,
\end{equation}
a shift in $a$, which should not change any physical observable, induces a shift in $\theta$. Another way of seeing this is by looking at how $Q_a$ acts on a given $\theta$ vacuum of the pure gauge sector (in the absence $a$, there is no shift symmetry and $\theta$ is observable). Under a gauge transformation in the homotopy class $n=1$, we have
\begin{equation}
    |\theta\rangle \to e^{-i\theta}|\theta \rangle.
\end{equation}
So, since $Q_a$ transforms as Eq.~\eqref{Qa_gauge_transf}, we have \cite{Jackiw:1976pf}
\begin{equation}\label{eq:vac_transform}
    \exp\left({\frac{i\theta'}{16\pi^2\lambda} Q_a}\right)|\theta\rangle = |\theta + \theta'\rangle.
\end{equation}
As $Q_a$ is conserved (commutes with the Hamiltonian), different theta vacua are degenerated in energy and thus physically indistinguishable. 

This degeneracy between the choice of $\theta$ and the zero mode (constant part) of $a$ has a different nature than the usual quantum symmetries. Consider, for instance, the symmetry breaking potential $V(\Phi) = \lambda_\Phi\left(|\Phi|^2 - \mu^2_\Phi/2\lambda_\Phi\right)^2$ for the complex scalar field $\Phi(x) = r(x) e^{i\alpha(x)}$. In this case, a vacuum expectation value for $r(x)$ spontaneously breaks the symmetry while $\alpha$ parametrizes the vacuum manifold, with shifts in $\alpha$ being the non-linear realizations of U$(1)$ phase rotation of the complex scalar field. If we think of $a$ as a Goldstone mode after some phase transition (as is the case in the Peccei-Quinn (PQ) proposal), then shifts around the vacuum manifold at constant $\theta$ are generated by the CS-independent part of $Q_a$, that we will call $\Tilde{Q}_a$:
\begin{equation}
    Q_a = \Tilde{Q}_a + Q_{\rm CS}, \quad \Tilde{Q}_a = \int dS \Dot{a}.
\end{equation}
Due to Eq.~\eqref{eq:vac_transform}, $\Tilde{Q}_{a}$ and $Q_{\rm CS}$ are not individually conserved, and so 
the conservation of $Q_a$ implies that $\Tilde{Q}_a$ is not conserved: shifts of the vacuum manifold for a given value of $\theta$ are not symmetries of the $\theta$ invariant quantum theory. Instanton effects explicitly break the continuous $\theta$ symmetry transformations. Therefore, there are no Goldstone bosons associated with this instanton induced explitict symmetry breaking. Note that it is a mistake to claim that the axion is a Golsdone boson associated with the anomalous breaking by these instanton effects. 

These arguments are at the core of 't Hooft's solution to the U$(1)$ problem in QCD \cite{Weinberg_PhysRevD.11.3583,tHooft:1986ooh}. The discussion about how the U$(1)$ symmetry is affected by non-perturbative effects is very similar to the chiral symmetry of massless fermions coupled to non-Abelian gauge fields. The standard chiral rotation combines with the instanton effects to make physics independent of $\theta$. Equivalently, the chiral rotation alone is explicitly broken by instantons. 

Returning to dCS gravity, we can apply the following non-Abelian axion results to the dCS pseudoscalar: ($i$) The dCS action has a continuous constant shift (quasi) symmetry with an associated divergent-free current $j^\mu_{\text{dCS}}$; ($ii$) \emph{If} gravity is quantized in the path integral for the theory, the CS-independent part of the total $\text{U}_\text{dCS}(1)$ charge is not conserved due to gravitational instanton effects\footnote{See \cite{Deser:1980kc} for a discussion on the gravitational theta sector.}; ($iii$) The total charge $Q_{\text{dCS}}$ \eqref{dCScharge} is conserved, but not invariant under homotopically non-trivial gauge transformations.  

So, we see that the U$(1)_{\text{dCS}}$ shift symmetry is always present, while the CS-independent part of the charge $\Tilde{Q}_{\text{dCS}}$ is only conserved in flat spacetime in the absence of gravitational instantons (which are absent at the semiclassical level). In that regime, $\Tilde{Q}_{\text{dCS}}$ is not conserved because the local coupling of $\phi$ with the gravitational Pontryagin density changes the dynamical evolution of the metric and the dCS scalar. For instance, in regions where $\Tilde{R}R$ is non-trivial, it will source $\phi$. This does not mean the symmetry under $\phi$ translations is broken: the total charge $Q_{\text{dCS}}$ is conserved (even after quantizing $\phi$). Similarly, one does not worry about non-trivial topological effects when using the axion coupling to $\Tilde{F}F$ to compute the axion-photon interactions or birefringence effects \cite{Adams:2022pbo, Blinov:2022tfy,Eskilt:2022cff}. 

In the next discussion, we shall couple a Dirac fermion with dCS and show that gravitational instantons do not affect a combination of the dCS-shift and fermion-axial currents.   

\section{Adding Fermions}\label{sec:coupling_with_fermions}

Before addressing SM fermions, we first  consider a single Dirac fermion $\psi$ in addition to the dCS action,
\begin{align}
    S[e^\mu_a,\phi,\psi] &= \int d^4 x \sqrt{-g}\left[\frac{1}{2\kappa^2}R -\frac{1}{2}\partial_\mu \phi \partial^\mu \phi - \right.\nonumber\\
    &\left.-\lambda \phi \Tilde{R}_{\mu\nu\rho\sigma}R^{\mu\nu\rho\sigma} - \Bar{\psi}e^{\mu}_a \gamma^a\left(\partial_\mu +\frac{1}{4} \omega^{ab}_\mu\gamma_{ab}\right)\psi\right],
\end{align}
where $e^\mu_a$ is the vierbein field and $\gamma_{ab} = \gamma_{[a}\gamma_{b]}$ are the Lorentz group generators for Dirac fermions. This classical theory has the global symmetry group $\text{U}(1)_{\text{dCS}}\times \text{U}(1)_{\text{V}} \times \text{U}(1)_{\text{A}}$. The new vector and axial $\text{U}(1)$'s are phase and chiral phase transformations of the fermion, $\psi \to e^{i\alpha}\psi$ and $\psi \to e^{i\beta \gamma_5}\psi$, respectively. Note that the fermion does not transform under $\phi$ shifts. The new Noether currents are the ordinary vector and axial currents,
\begin{equation}
    j^\mu_{\text{V}} = i\Bar{\psi} \gamma^\mu \psi, \quad j^{\mu}_{\text{A}} = i\Bar{\psi} \gamma^\mu \gamma_5 \psi,
\end{equation}
which are classically conserved upon using the equations of motion of the fermion. 

If we consider this theory at the quantum level, the currents will satisfy
\begin{subequations}
\begin{align}\label{gravanomaly}
    \nabla_\mu\left(\langle j^\mu_{\text{A}} \rangle - \frac{1}{192\pi^2}K^\mu\right) &= 0,\\
    \nabla_\mu\left(\langle \Tilde{j}^\mu_{\text{dCS}}\rangle + \lambda K^\mu \right) &=0,\\
    \nabla_\mu \langle j_\text{V}^\mu \rangle &=0.
\end{align}
\end{subequations}

The appearance of $K^\mu$ in the divergence of $j^\mu_\text{A}$ is due to a non-trivial path-integral Jacobian under axial transformations. Given its dependence on the curvature tensor, which has the same structure as the $\phi$-curvature coupling in dCS, one might wonder if a chiral rotation can modify the value of $\lambda$. In other words, there is a linear combination of the scalar shift and chiral rotation that is not affected by instanton effects at the quantum level, and one might wonder whether we can use such a combined U$(1)$ to hide the dCS scalar coupling inside the fermion chiral phase.  If so, the pseudoscalar-Pontryagin coupling in dCS gravity would not be observable. This is not the case, as we shall see below, since a constant chiral rotation cannot make a whole dynamic coupling disappear. 

We write the path integral for the quantum theory as:
\begin{widetext}
\begin{equation}
    Z[e, \omega] = e^{iS_{\text{EH}}}\int [d\phi d\Bar{\psi} d\psi] \exp \left\{i\int d^4 x \sqrt{-g}\left[-\frac{1}{2}\partial_\mu \phi \partial^\mu \phi - \lambda \phi \Tilde{R}_{\mu\nu\rho\sigma}R^{\mu\nu\rho\sigma} - \Bar{\psi}e^{\mu}_a \gamma^a\left(\partial_\mu +\frac{1}{4} \omega^{ab}_\mu\gamma_{ab}\right)\psi\right]\right\},
\end{equation}
\end{widetext}
where we brought the Einstein-Hilbert action $S_{\text{EH}}$ contribution to outside the path integral. Under the field redefinition
\begin{equation}\label{psiredefinition}
    \psi(x) \to \Tilde{\psi}(x) = e^{i\beta(x) \gamma_5} \psi(x), \; \Bar{\psi}(x) \to \Tilde{\Bar{\psi}}=\Bar{\psi}(x)e^{i\beta(x) \gamma_5},
\end{equation}
we have $[d\Bar{\psi}d\psi] \to [d\Tilde{\Bar{\psi}}d\Tilde{\psi}] = [d\Bar{\psi}d\psi]J[\beta]$, where \cite{Alvarez-Gaume:1983ihn,Fujikawa:2004cx}
\begin{equation}
    J[\beta] = \exp\left[-i\int d^4x \sqrt{-g} \frac{\beta(x)}{192\pi^2}\Tilde{R}_{\mu\nu\rho\sigma}R^{\mu\nu\rho\sigma}\right].
\end{equation}
Together with the fact that $Z[e, \omega]$ does not change by a change of integration variables, this result implies Eq.~\eqref{gravanomaly}. That $Z[e, \omega]$ is invariant also explains why there is no choice of $\beta(x)$ in the Jacobian that kills the $\phi \Tilde{R}R$ term in the action, as can be seen from
\begin{widetext}
\begin{align}\label{pathintredefinition}
    Z[e, \omega] = e^{iS_{\text{EH}}}\int [d\phi d\tilde{\Bar{\psi}} d\tilde{\psi}] &\exp\left\{i\int d^4 x \sqrt{-g}\left[-\frac{1}{2}\partial_\mu \phi \partial^\mu \phi - \lambda \phi \Tilde{R}_{\mu\nu\rho\sigma}R^{\mu\nu\rho\sigma} - \tilde{\Bar{\psi}}e^{\mu}_a \gamma^a\left(\partial_\mu + \frac{1}{4}\omega^{ab}_\mu\gamma_{ab}\right)\tilde{\psi}\right]\right\} \nonumber\\
    = e^{iS_{\text{EH}}}\int [d\phi d\Bar{\psi} d\psi] &\exp\left\{i\int d^4 x \sqrt{-g}\left[-\frac{1}{2}\partial_\mu \phi \partial^\mu \phi - \left(\lambda \phi +\frac{\beta}{192\pi^2}\right) \Tilde{R}_{\mu\nu\rho\sigma}R^{\mu\nu\rho\sigma} - \Bar{\psi}e^{\mu}_a \gamma^a\left(\partial_\mu +\frac{1}{4} \omega^{ab}_\mu \gamma_{ab}\right)\psi-\right.\right.\nonumber\\
    &-\left.\left. \partial_\mu \beta j^\mu_\text{A}\right]\right\},
\end{align}
\end{widetext}
so, even if we choose  $\beta(x) =  -192 \pi^2 \lambda \phi$, we will find
\begin{widetext}
\begin{align}
    Z[e, \omega] &= e^{iS_{\text{EH}}}\int [d\phi d\Bar{\psi}d\psi]  \exp\left\{i\int d^4 x \sqrt{-g}\left[-\frac{1}{2}\partial_\mu \phi \partial^\mu \phi - \Bar{\psi}e^{\mu}_a \gamma^a\left(\partial_\mu +\frac{1}{4} \omega^{ab}_\mu\gamma_{ab}\right)\psi-192\pi^2 \lambda \phi \nabla_\mu j^\mu_\text{A}\right]\right\}\nonumber\\
    &= e^{iS_{\text{EH}}}\int [d\phi d\Bar{\psi}d\psi]  \exp\left\{i\int d^4 x \sqrt{-g}\left[-\frac{1}{2}\partial_\mu \phi \partial^\mu \phi - \Bar{\psi}e^{\mu}_a \gamma^a\left(\partial_\mu +\frac{1}{4} \omega^{ab}_\mu\gamma_{ab}\right)\psi-\lambda  \phi \Tilde{R}_{\mu\nu\rho\sigma}R^{\mu\nu\rho\sigma}\right]\right\},
\end{align}
\end{widetext}
where to get the second equality we used the fact that, inside the path integral, the axial current satisfies Eq.~\eqref{gravanomaly}. So, we cannot simply set $\lambda$ to zero, and the $\phi \tilde{R} R$ coupling of dCS is quantum mechanically non-trivial even after the coupling with fermions. Equivalently, we could have redefined $\phi \to \phi - (192\pi^2\lambda)^{-1}\beta$ in the last line of Eq.~\eqref{pathintredefinition}, keeping the $\Tilde{R}R$ coupling.

One redefinition that can kill some part of $\phi$, however, is to assume a constant $\beta$. In this case, the last term inside the exponent of the last line in Eq.~\eqref{pathintredefinition} will not contribute, and thus, we cannot recover a $\tilde{R}R$ term from it. This is because a field redefinition of the form Eq.~\eqref{psiredefinition} with a constant $\beta$ is a genuine global transformation. Thus, quantum mechanically, the (constant) zero mode $\phi_0$ of $\phi(x^\mu)$ can be absorbed into the chiral phase of the fermions. Although such a phase is not observable, phase differences are, so $\lambda$ might still be measurable across regions where $\phi_0$ changes abruptly (e.g., domain wall solutions). More interestingly, if $\psi$ has a mass, only a combination of $\phi_0$ and the mass phase will be observable, and any measurement or observation sensible to the phase of the mass and $\phi_0$ will also constrain the value of $\lambda$. Note that the mass term's explicit chiral symmetry breaking will not affect these conclusions. On top of that, the results above are naturally extended for couplings with multiple fermions. 

The results in this section are a direct consequence of the fact that the path integral is invariant under redefinitions of the integrating fields. Our goal here was to explicitly check that if one wants to use the nontrivial Jacobian (anomaly) to remove the CS coupling with $\phi$ then, by consistency, one cannot neglect that $\nabla_\mu j^\mu_\text{A}\neq 0$, which ends up reintroducing the CS coupling. The same calculations also allow us to conclude that, when there is a mass term for the fermions, $\phi_0$ alone is not physical since only a combination of zero mode of $\phi$ and the mass matrix phase is observable.


\section{A UV completion of a dCS sector in the SM}\label{sec:coupling_with_SM}

In this section, we will propose a renormalizable four-dimensional model in which a dCS sector appears after integrating right-handed neutrinos $\Psi$ in an extension of the SM. We start by identifying the dCS scalar as the phase of a complex scalar whose Yukawa coupling with $\Psi$ gives them a Majorana mass after spontaneous symmetry breaking. We then couple the sterile right-handed neutrinos with the left-handed neutrinos via another Yukawa coupling and deduce the EFT for the latter and the dCS pseudoscalar below the symmetry-breaking scale.

\subsection{DCS gravity from integrating out heavy fermions}

Similar to axions arising from a PQ symmetry breaking, we can realize the dCS pseudoscalar as a Goldstone boson associated with the spontaneous breaking of a U$(1)$ symmetry. The remaining constant scalar shift is then a non-linear realization of the global symmetry below the breaking scale. We review this picture in the following model, recently discussed in Ref.~\cite{Alexander:2022cow} (see also Ref.~\cite{Bonnefoy:2020gyh} for the non-Abelian axion case), 
\begin{align}\label{dCS_UVaction}
    S = \int d^4 x \sqrt{-g}&\left[- \Bar{\Psi}e^\mu_a \gamma^a\left(\partial_\mu +\frac{1}{4} \omega^{cd}_\mu\gamma_{cd}\right)\Psi - |\partial \Phi|^2  \right.\nonumber\\
    &\left. -y\Bar{\Psi}\left(\Phi P_R + \Bar{\Phi}P_L\right)\Psi- V(|\Phi|) \right],
\end{align}
where $P_{L,R} = (1\pm \gamma_5)/2$ are the chiral projectors and $V(|\Phi|)$ is a symmetry-breaking potential for the complex field $\Phi$ whose explicit form is not important in the following. The action in Eq.~\eqref{dCS_UVaction} is invariant under the global transformation
\begin{align}\label{UV_global_sym}
    \Psi \to e^{i\beta \gamma_5}\Psi, \quad \Bar{\Psi} \to \Bar{\Psi}e^{i\beta \gamma_5}, \quad \Phi \to e^{-2i\beta}\Phi.
\end{align}
However, the associated path integral is not invariant as the symmetry is anomalous at the quantum level due to the gravitational contribution to the chiral anomaly. This symmetry is spontaneously broken by the vacuum expectation value of $\Phi$. Writing $\Phi(x^\mu) = (1/\sqrt{2})[v + \rho(x^\mu)]\exp\left[{i\phi(x^\mu)/v}\right]$, with $v$ a constant, we obtain
\begin{align}\label{symbrokenaction}
    S = \int d^4 x \sqrt{-g}&\left[- \Bar{\Psi}e^\mu_a \gamma^a\left(\partial_\mu +\frac{1}{4} \omega^{cd}_\mu\gamma_{cd}\right)\Psi \right.  \\
    &\left.- \frac{1}{2}(\partial \rho)^2 - \frac{1}{2}(1+\rho/v)^2(\partial \phi)^2 \right. \nonumber\\
    &\left. - \frac{y}{\sqrt{2}}(v+\rho)\Bar{\Psi}\left(e^{i \phi/v} P_R +  e^{-i\phi/v}P_L\right)\Psi - V(\rho)\right],\nonumber 
\end{align}
where $\phi$ is massless as expected from a Goldstone mode. In order to obtain dCS gravity, we now integrate out the fermion $\Psi$ and massive scalar $\rho$. In doing so, we obtain an effective field theory (EFT) for the pseudoscalar $\phi$ below the symmetry-breaking scale,
\begin{widetext}
\begin{equation}\label{psi_pathintegral}
    Z[\phi] = \int [d\Bar{\Psi} d\Psi d\rho] \exp\left\{i\int d^4x \sqrt{-g}\left[- \Bar{\Psi} \left(e^\mu_a \gamma^a\nabla_\mu +\frac{y}{\sqrt{2}}(v+\rho)e^{i(\phi/v)\gamma_5}\right)\Psi -\frac{1}{2}(\partial \rho)^2 - \frac{1}{2}(1+\rho/v)^2(\partial \phi)^2 -  V(\rho)\right]\right\},
\end{equation}
\end{widetext}
where we simplified the Yukawa coupling by rearranging the phases times chiral projectors into $\exp\left[{i(\phi/v)\gamma_5}\right]$. Using the field redefinition
\begin{equation}\label{chiral_redefinition}
    \Psi \to e^{-i\frac{\phi}{2v}\gamma_5}\Psi, \quad \Bar{\Psi} \to \Bar{\Psi}e^{-i\frac{\phi}{2v}\gamma_5}, 
\end{equation}
we get
\begin{widetext}
\begin{align}\label{symbrokenPI}
    Z[\phi] = \int [d\Bar{\Psi} d\Psi d\rho] &\exp\left\{i\int d^4x \sqrt{-g}\left[- \Bar{\Psi} \left(e^\mu_a \gamma^a\nabla_\mu +\frac{y}{\sqrt{2}}(v+\rho) - \frac{i}{2v}\partial_\mu \phi \gamma^\mu \gamma_5 \right)\Psi -\frac{1}{2}(\partial \rho)^2 - \frac{1}{2}(1+\rho/v)^2(\partial \phi)^2 -  \right.\right.\nonumber\\
    &+\left.\left. V(\rho)+ \frac{1}{192\pi^2}\frac{\phi}{2v}R\Tilde{R}\right]\right\},
\end{align}
\end{widetext}
where the last term in the exponential appears due to the non-trivial Jacobian of the path integral measure under the change of variables above and we identified the spinor-covariant derivative $\nabla_\mu \Psi = (\partial_\mu + \omega_\mu^{ab}\gamma_{ab}/4)\Psi$. To leading order, the path integral picks the classical saddle where $\Psi = 0 =\rho$,  so the path integral then becomes 
\begin{equation}\label{lowEaction1}
    Z[\phi] \propto \exp\left\{{i\int d^4x \sqrt{-g}\left[-\frac{1}{2}(\partial \phi)^2 + \frac{\phi}{384\pi^2v} R\Tilde{R} + \ldots \right]}\right\},
\end{equation}
where dots stand for higher-derivative corrections suppressed by the fermionic and $\rho$ masses.\footnote{Since the fermionic path integral was done in curved space, there will be also extra  curvature contributions, see Refs.~\cite{Mauro:2014eda,Toms:2018wpy} and references therein.} We see that the dCS gravity term is obtained from integrating-out heavy fermions. This model is thus analogous to the Kim-Shifman-Vainshtein-Zakharov (KSVZ) model for the QCD axion \cite{Kim:1979if, Shifman:1979if}. Again, the shift symmetry of the dCS pseudoscalar is a non-linear realization of the non-anomalous phase transformation of $\Phi$ in Eq.~\eqref{UV_global_sym}. It is worth noting that the theories defined the action in Eq.~\eqref{symbrokenaction}  and Eq.~\eqref{symbrokenPI} are dual since their path integrals are the same (they differ by a fermionic $\Psi$ field redefinition), with the latter having no $\lambda \phi R\Tilde{R}$ coupling. However, once we integrate out $\Psi$, there is no fermionic field redefinition left, and the $\lambda \phi R \Tilde{R}$ coupling is a genuine low-energy operator of the effective action in Eq.~\eqref{lowEaction1}. 

\subsection{Coupling a DCS sector to the SM}

In what follows, we wish to make a connection with the Standard Model where the neutrinos get a Majorana mass term at energies below the symmetry-breaking scale, with $\phi$ appearing as a phase in such a mass term. We do so after assuming that the heavy fermion $\Psi$ plays the role of a right-handed sterile neutrino that couples with the lepton and Higgs doublets, $L$ and $H$, respectively,
\begin{equation}
    L = \begin{pmatrix}
        L_\nu \\
        L_e
    \end{pmatrix}, \quad \tilde{H} = \begin{pmatrix}
        {\phi^0}^* \\
        -{\phi^+}^*
    \end{pmatrix},
\end{equation}
as 
\begin{equation}\label{YukawaRnu}
    \Delta \mathcal{L} = -\Tilde{y}(\Bar{L} \tilde{H} P_R \Psi + \Bar{\Psi}P_L \Tilde{H}^\dagger L),
\end{equation}
where $L_\nu$, $L_e$, and $\Psi$ are Majorana spinors:
\begin{equation}
    L_f = \begin{pmatrix}
        f_L \\
        i\sigma^2 f_L^*
    \end{pmatrix}, \quad \Psi = \begin{pmatrix}
        -i\sigma^2 \nu_R^* \\
        \nu_R
    \end{pmatrix}.
\end{equation}
Note that we need the conjugated $\Tilde{H}$  instead of the Higgs doublet $H^T = (\phi^0 \;\; \phi^+)$ in Eq.~\eqref{YukawaRnu} to get a hypercharge neutral coupling. We consider a single lepton flavor for simplicity since a generalization for all three generations is straightforward (see the discussion section).

With the new Yukawa coupling \eqref{YukawaRnu} and the kinetic terms for $L$ and $H$, the action in Eq.~\eqref{dCS_UVaction} becomes
\begin{align}\label{dCS_UVaction2}
    S = \int d^4 x \sqrt{-g}&\left[-\frac{1}{2}\Bar{\Psi}\slashed \nabla \Psi - \partial_\mu \Phi \partial^\mu \Bar{\Phi} - \frac{1}{2}\Bar{L} \slashed \nabla L - \right.\nonumber\\
    &\left. -|\partial H|^2 -y\Bar{\Psi}\left(\Phi P_R + \Bar{\Phi}P_L\right)\Psi-  \right.\nonumber\\
    &\left. -\Tilde{y}(\Bar{L} \tilde{H} P_R \Psi + \Bar{\Psi}P_L \Tilde{H}^\dagger L)- V(|\Phi|) \right].
\end{align}
This action is invariant under Eq.~\eqref{UV_global_sym} provided $L$ transforms as 
\begin{align}
    L\to e^{i\beta \gamma_5}L. \label{UV_global_sym2}
\end{align}
Note that since both $L$ and $\psi$ are four-component Majorana spinors, the global chiral transformation acting on those is a phase rotation associated with their fermionic number. We discuss further aspects of this symmetry in the discussion section. After performing the field redefinition in Eq.~\eqref{chiral_redefinition}, the path integral is now
\begin{widetext}
\begin{align}
    Z = \int [d\Bar{\psi} d\psi da] \text{exp}&\left\{i\int d^4x \sqrt{-g}\left[- \frac{1}{2}\Bar{\Psi} \left(e^\mu_a \gamma^a\nabla_\mu +\frac{y}{\sqrt{2}}(v+\rho) - \frac{i}{2v}\partial_\mu \phi \gamma^\mu \gamma_5 \right)\Psi -\frac{1}{2}(\partial \rho)^2 - \right.\right.\nonumber\\
    &\left.-|\partial H|^2 - \Bar{L}e^\mu_a \gamma^a \nabla_\mu L -\frac{1}{2}(1+\rho/v)^2(\partial \phi)^2 -V(\rho)+\frac{1}{192\pi^2}\frac{\phi}{2v}R\Tilde{R}-\right.\nonumber\\
    &\left.\left.-\tilde{y}\left(\Bar{L}\Tilde{H}e^{-i\frac{\phi}{2v}\gamma_5}P_R\Psi + \Bar{\Psi}P_Le^{-i\frac{\phi}{2v}\gamma_5}\Tilde{H}^\dagger L\right)\right]\right\}.
\end{align}
\end{widetext}
We thus see that $\bar{\chi} = -\Tilde{y} \Bar{L}\Tilde{H} e^{-i\frac{\phi}{2v}\gamma_5}$ acts like a source for the right-hand part of $\Psi$. Moreover, using $\Bar{L} = L^c=L^{T}C$, where $C$ is the charge conjugation matrix, we have
\begin{equation}
     \Bar{\Psi}P_Le^{-i\frac{\phi}{2v}\gamma_5}\Tilde{H}^\dagger L = - \Psi^T P_L e^{-i\frac{\phi}{2v}\gamma_5}\Tilde{H}^\dagger \Bar{L}^{T}
\end{equation}
and so $\eta = e^{-i\frac{\phi}{2v}\gamma_5}\Tilde{H}^\dagger L$ acts like a source for the left-hand part of $\Psi$. But since $\Psi$ is a Majorana spinor, its left and right components are related, and both $\chi$ and $\eta$ contribute as sources to the path integral of $\Psi$. Hence, the coupling in Eq.~\eqref{YukawaRnu} will give the contribution 
\begin{align}
    -\frac{1}{2}\Bar{\chi}(\slashed D - M)^{-1}P_R\chi -\frac{1}{2}\Bar{\eta}(\slashed D - M)^{-1}P_L\eta \nonumber\\
    = -\frac{1}{2m_\Psi}\left(\Bar{\chi}P_R\chi+\Bar{\eta}P_L \eta\right) +\cdots
\end{align}
to the effective Lagrangian. Note that the charge conjugation of the chiral sources fixes $\chi$ and $\eta$, 
\begin{align}
    \chi &\equiv C \Bar{\chi}^T = -\Tilde{y}e^{-i\frac{\phi}{2v}\gamma_5}\Tilde{H}^T L, \\
    \Bar{\eta} &\equiv \eta^T C= \Bar{L}\Tilde{H}^* e^{-i\frac{\phi}{2v}\gamma_5}.  
\end{align}
The effective action is given by
\begin{align}\label{final_effective_action}
    S &= \int d^4x \sqrt{-g}\left[-\frac{1}{2}(\partial \phi)^2 -|\partial H|^2 -\frac{1}{2}\Bar{L}e^\mu_a \gamma^a \nabla_\mu L -\right.\nonumber\\
    &-\left.\frac{\Tilde{y}^2}{2m_\Psi}\left(e^{i\frac{\phi}{v}}\Bar{L}\Tilde{H} P_R \Tilde{H}^T L + e^{-i\frac{\phi}{v}}\Bar{L}\Tilde{H}^* P_L \Tilde{H}^\dagger L\right) -\right.\nonumber\\
    &\left.-\frac{\phi}{384\pi^2 v} R\Tilde{R} + \cdots \right],
\end{align}
where the heavy fermion mass $m_\Psi$ can be read from Eq.~\eqref{psi_pathintegral} to be $m_\Psi = yv/\sqrt{2}$. The resulting effective action has a modified Weinberg operator \cite{Weinberg:1979sa} that includes a coupling with $\phi$.

Notice that the pseudoscalar arises as a complex phase in the neutrino mass matrix and as a coupling to the gravitational Pontryagin density. A measurement of the mass phase will not constrain its coupling to the gravitational Pontryagin density because strong gravitational effects from compact binary can source a large gravitational Pontryagin density which will source large gradients of $\phi$. 

This is analog to QCD, as the axion couples to chromo-electric and magnetic field via the Pontryagin density, astrophysical sources may enhance the axion and a similar story applies to dCS.  Here, there are two observable effects to consider.  First, an Earth-based measurement of the dynamical field can constrain the value of the pseudoscalar, which is predicted to be small due to the fact that the Pontryagin term is zero in Minkowski space-time. However in compact binary systems strong gravity can give a non-vanishing $R\tilde{R}$ which will in turn source gravitational waves and non-vanishing pseudoscalar amplitude.  Also, in models of Higgs-Inflation, Electroweak symmetry is broken during inflation, and the phase can also be non-vanishing in the early universe.  In these strong gravity, early universe regimes the mass matrix phase will also be non-vanishing but obviously difficult to measure. Still, it may be possible to study the effects of the phase and its impact on neutrino cosmology. We leave this possibility for future investigation.

\section{Discussion and Conclusion}\label{sec:conclusion}

In this paper, we have discussed many symmetry aspects of dCS gravity and anomalies that are relevant when trying to get the dCS scalar-gravity coupling from a UV theory. Most of the necessary concepts and results have analogs in the physics of axions and their coupling with gauge fields. We reviewed some known results on axions in detail to better understand their relation and applicability to the gravitational case. We also proposed an extension of the SM model that gives origin to a dCS sector and the coupling of the dCS pseudoscalar with neutrinos at low energies. In the rest of this section, we discuss the model's symmetry and its gauging.

The modified Weinberg operator in Eq.~\eqref{final_effective_action} breaks the lepton number symmetry (as the SM lepton number symmetry alone is anomalous, we consider the baryonic minus lepton number symmetry). But there is a combination of the U$(1)_{\text{B}-\text{L}}$ and the $\phi$ shift symmetry that leaves the modified Weinberg term invariant and is a quasi-symmetry of the action Eq.~\eqref{final_effective_action}. However, such a symmetry is anomalous and the $\phi R\Tilde{R}$ coupling cannot be eliminated by redefining $L$ in the path integral for the same reasons explained in Section \ref{sec:coupling_with_fermions}: such a redefinition would bring a term proportional to the divergence of the $L$'s chiral current (after integration by parts), giving back the dCS coupling. We comment on how the symmetry combination above can be gauged from the low-energy effective perspective in Appendix \ref{SM_anomalies}.

On the other hand, the action in Eq.~\eqref{dCS_UVaction2} is invariant under the simultaneous rotation of $\Psi$, $L$, and $\Phi$ (with appropriate charges) Eq.~\eqref{UV_global_sym} and Eq.~\eqref{UV_global_sym2}, and such symmetry is non-anomalous because the gravitational contribution of $L$ and $\Psi$ to the anomaly cancel each other. Hence, we can gauge this non-anomalous symmetry combination, which we call U$(1)_{\text{B}-\text{L}-\phi}$. This begs the question of what happens with this gauge symmetry after integrating $\Psi$. At energies well below the $\Psi$ mass, the only fermion left in the spectrum is the chiral $L$, which contributes to the anomaly of the gauge symmetry. As discussed in Ref.~\cite{Preskill:1990fr}, despite this anomaly, this effective chiral gauge theory is unitary (but not renormalizable) because the gauge boson acquires its mass via a Higgs mechanism: the gauge symmetry is non-linearly realized as a shift symmetry of the gauge boson longitudinal mode. The main consequence of this fact is that $\phi$ appears in the EFT as the longitudinal mode of the gauge boson in a Stueckelberg $(A^\mu, \phi)$ sector and hence gets eaten by the gauge field in the unitary gauge. Moreover, before integrating $\Psi$, the Lagrangian includes the term $A_\mu j^\mu_\Psi = A_\mu i\Bar{\Psi}\gamma^\mu\gamma_5 \Psi$ and the gauge transformation of the effective action necessary to select the unitary gauge will give rise to a term proportional to $\phi \partial_\mu j^\mu_\Psi$, which after the $\Psi$ path-integration will appear in the effective Lagrangian as $\phi(R\Tilde{R} - F\Tilde{F})$. The gauge variation of this term cancels the non-trivial Jacobian from the $L$'s fermionic measure, making the effective action gauge invariant at the quantum level.

The results in the previous sections can be straightforwardly generalized to an arbitrary number of heavy fermions. If we consider the case of all left-handed SM leptons $L_i$ and three sterile right-handed neutrinos $\Psi_i$, we have the simplest Majoron model \cite{Chikashige:1980ui, Chikashige:1980qk}, where $\phi$ play the role of a Majoron\footnote{See Refs.~\cite{Latosinski:2012zgs, Latosinski:2012qj} for a proposal of identifying the QCD axion with the Majoron.}. The $y$ and $\Tilde{y}$ Yukawa couplings will be promoted to matrices $y_{ij}$ and $\Tilde{y}_{ij}$ with flavor indices and the Weinberg operator in the effective Lagrangian would read
\begin{equation}
    -\tilde{y}_{im}M^{-1}_{mn}\tilde{y}_{nj}\left(e^{i\frac{\phi}{v}}\Bar{L_i}\Tilde{H} P_R \Tilde{H}^T L_j + e^{-i\frac{\phi}{v}}\Bar{L}_i\Tilde{H}^* P_L \Tilde{H}^\dagger L_j\right),
\end{equation}
withe $M_{mn}^{-1} = y_{mn}v/\sqrt{2}$. An interesting question is how $\phi$ affects the seesaw mechanism, a topic that we leave for future investigations.

The Majoron model above has the global non-anomalous symmetry U$(1)_{\text{B}-\text{L}-\phi}$ that, if gauged, would hide the dCS scalar $\phi$ in the longitudinal mode of a massive gauge boson\footnote{For gauged Majoron models, see Refs.~\cite{Rothstein:1992rh, Ibe:2018hir}}. Note that if we start with a different number of heavy fermions, unmatched with the three left-handed neutrinos, we cannot straightforwardly gauge U$(1)_{\text{B}-\text{L}-\phi}$ because this symmetry will be anomalous from the on-set. Moreover, even for three right-handed neutrinos, gauging the symmetry is a choice\footnote{We thank Michael Peskin for bringing up this point to our attention.} and there is nothing inconsistent with having a global, non-anomalous, U$(1)_{\text{B}-\text{L}-\phi}$. In this case, $\phi$ is a dynamical field that couples with the Pontryagin density. This is analogous to the PQ symmetry and axions: although there are models where U$(1)_\text{PQ}$ is gauged and the axion is eaten to give the gauge boson a mass, this has not precluded the search for axions in particle physics, astrophysics and cosmology.

Although there are motivations for gauging global symmetries from the string theory perspective, the same point of view also motivates us to rethink the origin of the dCS: we know that a dCS sector should appear in \emph{any} heterotic four-dimensional compactification due to the Green-Schwarz mechanism (see discussion in Ref.~\cite{Alexander:2009tp}). This is because $\phi$ can be identified with the zero mode of the ten-dimensional Kalb-Rammond field on the internal space, and this identification is compactification-independent ($\phi$ is referred to as the model-independent axion) \cite{Svrcek:2006yi}. Then, in the most basic example of how to get dCS from string theory, the way the global dCS shift symmetry becomes local is by a promotion to the gauge symmetry of a 2-form at energies close to and larger than the compactification scale. Thus, at least from a theoretical perspective, a dCS sector can appear at low energies without any inconsistency between its global symmetry and quantum gravity. More realistically, since $\phi$ pairs with the dilaton, it gets a potential (generated by non-perturbative effects, see Ref.~\cite{Cicoli:2013rwa} for a review) that breaks its continuous global symmetry down to a discrete group, again in analogy to the PQ axion, and so it cannot be the longitudinal mode of any gauge field.




\begin{acknowledgments}
We thank Chris Hull for enlightening discussions, Peter Adshead, Robert Brandenberger, Keshav Dasgupta, Jim Gates, Greg Gabadadze, Humberto Gilmer, Michael Peskin, Sanjaye Ramgoolam, and Matt Strassler for discussion and feedback on a draft of this paper. 
\end{acknowledgments}

\appendix

\section{Baryon and Lepton Number Anomalies in the SM}\label{SM_anomalies}

In this appendix, we state some results on global anomalies in the Standard Model and comment on how the anomaly changes after introducing a dCS sector. The electroweak sector is a chiral gauge theory due to the chiral coupling of SU$(2)_{\text{L}}$ gauge bosons with quarks and leptons. As it is well known, chiral gauge theories have gauge anomalies which have to be cancelled, and such is the case in the standard model after summing over all representations and hypercharges of the fermions. Chiral gauge theories have also a fermion number global anomaly, and in the SM, the global baryonic and leptonic number symmetries are indeed anomalous. This is because under redefinitions of the form
\begin{subequations}
\begin{align}
    \psi_L(x) \to \tilde{\psi}_L(x) &= e^{i\alpha(x)}\psi_L(x), \\
    \psi_R(x) \to \tilde{\psi}_R(x) &= e^{i\alpha(x)}\psi_R(x),
\end{align}
\end{subequations}
the left and right contributions to the Jacobian of the path integral measure do not cancel when only one component (left or right) couple with the gauge field. So, the number current $j_n^\mu(x) = i\Bar{\psi}(x) \gamma^\mu \psi(x)$ will satisfy \cite{Fujikawa:2004cx}
\begin{align}
    \partial_\mu \langle j_n^\mu \rangle &=  \partial_\mu \left(\langle i\bar{\psi}_L \gamma^\mu \psi_L\rangle + \langle i\bar{\psi}_R \gamma^\mu \psi_R \rangle \right) \nonumber\\
    &=\pm\frac{\text{tr}(T^a T^b)}{16\pi^2}\Tilde{F}^{a}_{\mu\nu}F^{\mu\nu b},
\end{align}
where the negative (positive) sign case is for when only the left (right) chirality couples with the gauge field. The trace is over the representation of the $\psi$ chirality that contributes to the anomaly. In the SM, only the left-hand part of the quark and lepton field doublets $q$ and $l$ couple with SU$(2)_\text{L}$:
\begin{align}
    S_{\text{SM}} \supset \int d^4 x &\left[-\bar{q}_L^i\gamma_\mu\left(\partial_\mu + ig A^a_\mu \frac{\sigma^a}{2}\right)q_L^i -\right.\nonumber\\
    &\left.-\bar{l}_L^i\gamma_\mu \left(\partial_\mu + ig A^a_\mu \frac{\sigma^a}{2}\right)l^i_L \right]. 
\end{align}
The indices in the doublets are generation indices and the fundamental-representation indices are suppressed. For instance,
\begin{equation}
    l_L^1 = \begin{pmatrix}
        \nu_e \\
        e_L
    \end{pmatrix}, \quad q_L^1 = \begin{pmatrix}
        u_L \\
        d_L
    \end{pmatrix}.
\end{equation}

The SM baryon number current satisfies
\begin{equation}
    \partial_\mu \langle j_\text{B}^\mu \rangle = \frac{3 N_c}{32\pi^2} \tilde{F}_{\mu\nu}^a F^{\mu\nu a},
\end{equation}
where $N_c$ is the number of quark colors. The lepton number current will also be anomalous
\begin{equation}
    \partial_\mu \langle j_\text{L}^\mu \rangle = \frac{3}{32\pi^2} \tilde{F}_{\mu\nu}^a F^{\mu\nu a}.
\end{equation}
The factor of $3$ in the numerator of the anomalies above comes from the contributions of the $3$ quark and lepton generations. This tells us that the baryon and lepton numbers are not conserved and, in particular, their associated U$(1)$ transformations cannot be straightforwardly gauged. 

However, the combination $j_{\text{B}-\text{L}}^{\mu}= j_\text{B}^\mu/3 - j^\mu_\text{L}$ satisfies
\begin{equation}
        \partial_\mu \langle j^{\mu}_{\text{B}-\text{L}}\rangle =  \frac{N_c -3}{32\pi^2} \tilde{F}_{\mu\nu}^a F^{\mu\nu a},
\end{equation}
and so is conserved in the SM, since $N_c= 3$ (note that the factor of $1/3$ in the baryonic contribution to $j_{\text{B}-\text{L}}^{\mu}$ is because quarks have a baryon number $B=1/3$). So, the Standard Model has a potential non-anomalous global U$(1)_{\text{B}-\text{L}}$ symmetry. In beyond SM models, one can add extra breaking effects for U$(1)_\text{B}$ and U$(1)_\text{L}$ individually, but the U$(1)_{\text{B}-\text{L}}$ should be non-anomalous. One can even contemplate gauging such a symmetry. 

The discussion above is modified when we consider the Standard Model in curved spacetimes. Since there are no right-handed neutrinos, there is a mismatch in the number of left and right chiralities in the fermionic sector. Recall that a single chiral fermion contributes to the gravitational anomaly of an otherwise non-anomalous current as \cite{Alvarez-Gaume:1983ihn}
\begin{equation}
    \nabla_\mu \langle i\Bar{\psi}_{L,R} \gamma^\mu \psi_{L,R}\rangle = \pm \frac{1}{384\pi^2}\Tilde{R}_{\mu\nu\rho\sigma} R^{\mu\nu\rho\sigma}, 
\end{equation}
where the negative (positive) sign appears for the left (right) chirality contribution. So, for a non-chiral fermion with propagating left and right-parts, we have
\begin{subequations}
\begin{align}
    \nabla_\mu \langle i\Bar{\psi} \gamma^\mu \psi\rangle &= \nabla_\mu \left[\langle i\Bar{\psi}_L \gamma^\mu \psi_L \rangle + \langle i\Bar{\psi}_R \gamma^\mu \psi_R \rangle \right]=0, \\
    \nabla_\mu \langle i\Bar{\psi} \gamma^\mu \gamma^5 \psi\rangle &= \nabla_\mu \left[\langle i\Bar{\psi}_L \gamma^\mu \psi_L \rangle - \langle i\Bar{\psi}_R \gamma^\mu \psi_R \rangle \right]\nonumber\\
    &= \frac{1}{192\pi^2}\Tilde{R}_{\mu\nu\rho\sigma} R^{\mu\nu\rho\sigma}.
\end{align}
\end{subequations}

Let us apply these results to $j^\mu_{\text{B}-\text{L}}$ in the SM. The left-handed contribution to the lepton number overcomes the right-handed one, such that the total lepton number current satisfies
\begin{equation}
    \nabla_\mu \langle j^\mu_L \rangle = \frac{3}{32\pi^2} \tilde{F}_{\mu\nu}^a F^{\mu\nu a} + \frac{3}{384\pi^2}\Tilde{R}_{\mu\nu\rho\sigma} R^{\mu\nu\rho\sigma}.
\end{equation}
where the factor of $3$ in the numerator of the gravitational contribution is the excess of left over right neutrinos. However, since the number of left and right-handed quarks is the same, there is no gravitational contribution to the $j_\text{B}^\mu$ anomaly. Hence
\begin{equation}
    \nabla_\mu \langle j^\mu_{B-L} \rangle =  \frac{3}{384\pi^2}\Tilde{R}_{\mu\nu\rho\sigma} R^{\mu\nu\rho\sigma}.
\end{equation}
In other words, since the lepton number current anomaly has a gravitational contribution while the baryonic current does not possess one, the gravitational contribution to the anomaly in $j^\mu_{\text{B}-\text{L}}$ cannot be canceled.  

Now, after introducing the dCS sector, following the discussion of Section \ref{sec:coupling_with_fermions}, there will be a combination of U$(1)_{\text{dCS}}$ and U$(1)_{\text{B}-\text{L}}$ that is not affected by instanton effects at the quantum level, i.e., it is non-anomalous. We will call this combination U$(1)_{\text{B}-\text{L}-\phi}$. Then, a constant part of $\phi$ can always be absorbed in the phase of the fermions, and only a combination of such a zero mode and the phase of the mass matrix determinant is observable. This is very similar to the CP violation in QCD with the zero mode of $\phi$ playing the role of the QCD $\theta$. 

In some extensions of the Standard Model, the U$(1)_{\text{B}-\text{L}}$ is ultimately gauged, leading to new effects. While this can be done in flat space, the gravitational contribution to the chiral anomaly must be consistently canceled for the gauging in curved spacetimes. However, with an added dCS sector, the non-anomalous U$(1)_{\text{B}-\text{L}-\phi}$ can seemingly be gauged without any constraints on the background, which looks attractive, for instance, if one wants to contemplate quantizing gravity. Such a procedure is subtle because despite U$(1)_{\text{B}-\text{L}-\phi}$ being non-anomalous at the quantum level, there is no contribution from the chiral anomaly to the conservation of the classical $j^\mu_{\text{B}-\text{L}-\phi}$ current. In other words, the partition function would be gauge invariant but not the classical action. Notwithstanding, global  anomalous symmetries can be gauged at the expense of non-renormalizability \cite{Preskill:1990fr,Harvey:1988in}, so the resulting gauge theory should be an EFT valid only below some cut off scale.


\bibliography{references}

\end{document}